# On-chip microwave-spin-plasmon interface (MSPI)

Mikhail Y. Shalaginov*, Simeon I. Bogdanov, Alexei S. Lagutchev, Alexander V. Kildishev, Alexandra Boltasseva, and Vladimir M. Shalaev*

*School of Electrical & Computer Engineering, Birck Nanotechnology Center, and Purdue Quantum Science & Engineering Institute, Purdue University, West Lafayette, IN 47907, USA*

*email: shalaginov@purdue.edu, shalaev@purdue.edu

## ABSTRACT

On-chip scalable integration represents a major challenge for practical quantum devices. One particular challenge is to implement on-chip optical readout of spins in diamond. This readout requires simultaneous application of optical and microwave fields along with an efficient collection of fluorescence. The readout is typically accomplished via bulk optics and macroscopic microwave transmission structures. We experimentally demonstrate an on-chip integrated structure for nitrogen-vacancy (NV) spin-based applications, implemented in a single material layer with one patterning step. A nanodiamond with multiple NV centres is positioned at the end of the groove waveguide milled in a thick gold film. The gold film carries the microwave control signal while the groove waveguide acts as a fluorescence collector, partially filtering out the pump excitation. As a result, the device dimensions and fabrication complexity are substantially reduced. Our approach will foster further development of ultra-compact nanoscale quantum sensors and quantum information processing devices on a monolithic platform. NV centre-based nanoscale sensors are the most promising application of the developed interface.

**Solid-state atomic impurities for integrated quantum applications**

Colour centres in solids are promising candidates for the development of integrated quantum devices. These defects possess unique optical and/or spin properties and are compatible with scalable on-chip engineering. Of particular interest are the quantum emitters realising the so-called spin-light interfaces[1]. Examples of such emitters include colour centres in diamond [2], silicon carbide[3], rare-earth ions[4], and donors in silicon[5]. The breakthrough applications of the solid-state defects encompass large-scale quantum simulators[6], and multifunctional nanoscale sensing[7], however, all of their demonstrations to date involved bulky table-top setups. There is a strong need for further miniaturisation and full on-chip integration of devices based on solid-state defects. The necessary components to be interfaced on-chip include pump sources, waveguides, the infrastructure for the control of electron spin states (e.g., microwave antennae and magnets), and photodetectors. Recently, several attempts have been undertaken towards CMOS-integrated quantum sensors with a multilayer architecture[8,9]. We demonstrate an approach that will allow to further scale down the device architecture and reduce fabrication complexity.

**Nitrogen-vacancy spin coupling to optical and microwave excitations**

In this work, we focus on nitrogen-vacancy (NV) centres in diamond[10], best known for their exceptional electron spin coherence at room temperature[11] and above[12]. NV's fluorescence intensity is spin-dependent, and this dependence is used for optical spin state readout. Upon optical excitation, the NV centre spontaneously decays to its ground orbital state following radiative or non-radiative paths (Fig. 1, level diagram). The probability of emitting a photon is dictated by the distribution of the centre's ground-state population over the spin-triplet sublevels $m_s = 0, \pm 1$. The non-radiative decay probability for the $m_s = \pm 1$ spin projections is noticeably higher, resulting in a reduced fluorescence signal compared to the $m_s = 0$ projection state. The relative difference in the fluorescence signals from these spin states is termed the spin contrast. The zero-field energy splitting between the $m_s = 0$ and $m_s = \pm 1$ orbital ground state sublevels is about 2.87 GHz, due to spin-spin interaction[13], and increases with the applied constant magnetic field by $\gamma_{NV} = 2.7$ MHz/G. The NV $m_s = 0 \rightarrow m_s = \pm 1$ transitions can be probed by applying an a.c. magnetic field while measuring the spin contrast. This contrast is observed when the microwave (MW) signal frequency is swept over the spin transition frequency. The electron spin resonance frequency is sensitive to not only the local d.c. magnetic field but also to electric field[14], temperature[15], pressure[16], and crystal strain[17], providing multifunctional nanoscale sensing capability. In addition, the electron spin resonance is essential for harnessing nearby nuclear spins to generate entangled spin states[18], long-lived quantum memories[19], and to perform quantum error correction[20].

The sensing functionalities of the NV centres rely on the readout of the NV spin population as a function of the external electromagnetic, strain or temperature fields. Several schemes for NV spin readout have been demonstrated in the past[21], including those based on fluorescence[22], spin-to-charge conversion[23], photocurrent[24], low-temperature resonance[25], and nuclear spin-assistance[26]. The fluorescence-based readout (later termed as optically detected magnetic resonance or ODMR) is advantageous due to its simplicity and room-temperature operation.

The sensitivity of spin population measurement in this scheme depends on the fluorescence rate $\gamma$, optical excitation rate $k_{exc}$, microwave Rabi frequency, setup efficiency, as well as on the achieved maximum spin contrast between the $m_s = 0$ and $m_s = \pm 1$ spin states. Generally speaking, one can increase the collected emission by enhancing the fluorescence rate $\gamma$, e.g., through Purcell effect while exciting an NV centre in the saturation regime ($k_{exc} > \gamma$), or by employing large NV ensembles (NVE). Here we focus on NVEs as they find practical applications in most sensing scenarios. In our prior work, we found that spin readout in dense NVEs is compromised in the optical saturation regime[27]. One is bound to operate in the weak excitation regime ($k_{exc} < \gamma$), in which the spin contrast is roughly inversely proportional to the fluorescence rate. Therefore, the fluorescence lifetime shortening must be restricted to moderate levels and the spin readout sensitivity may instead be enhanced via an efficient photon collection scheme. A suitable integrated fluorescence collecting structure for NVE spin readout is a waveguide featuring a high collection efficiency ($\beta$–factor) without a strong effect on the NVE fluorescence lifetime. So far, those conditions have been achieved in several architectures including diamond nanobeams [23,28], dielectric-loaded plasmonic waveguides [29], and channel plasmonic waveguides [30].

**Concept of on-chip microwave-spin-plasmon interface**

In the ODMR-based experiments, it is essential to deliver both optical and microwave excitations as well as collect the spin-dependent fluorescence. The delivery of the microwave excitation is usually performed using a dedicated conductor, separated from the optical pumping and fluorescence collection channels with the NV centre located within tens of micrometres of the wire/micro-antenna[28,31].

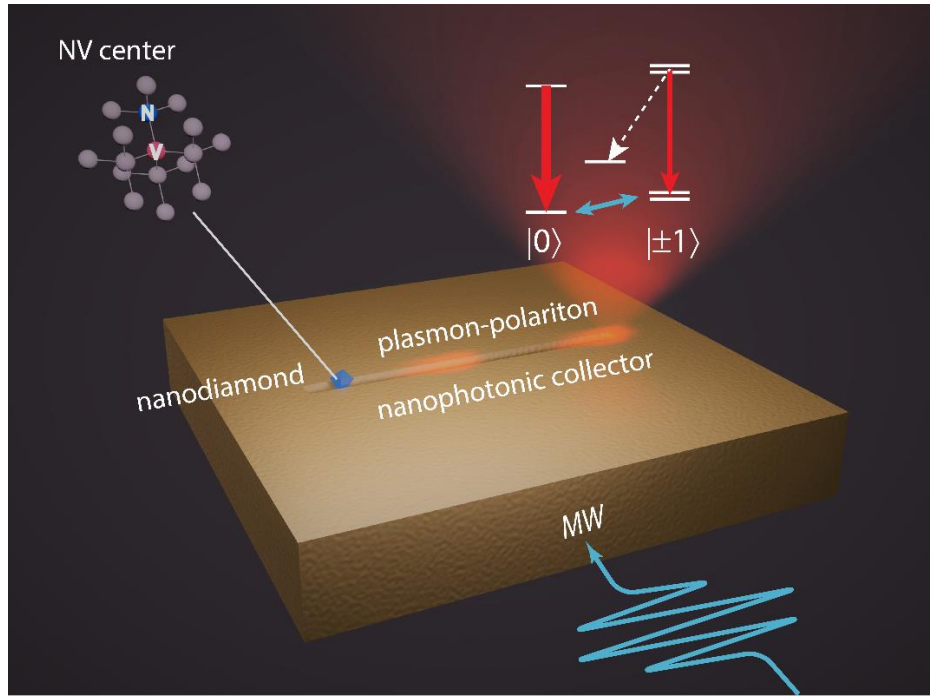

Fig. 1. Artistic rendition of the microwave-spin-plasmon interface (MSPI), for on-chip sensing and quantum information processing. The MSPI consists of a nanodiamond-based NVE coupled to a v-groove nanophotonic collector milled in a thick gold film. The NVE fluorescence is channelled into the nanophotonic collector and is outcoupled into the far field at its extremities. The gold film simultaneously acts as a carrier for the microwave excitation.

Simultaneous integration of MW and optical infrastructures can be achieved on a conducting platform, such as a thick metallic film patterned with nanophotonic collector-waveguides. It turns out that this structure can simultaneously deliver the MW control signal, couple the NV fluorescence into channel plasmon-polaritons[32] (termed as plasmons later in the text), and partially filter out the residual pump coupling into the collector. In this work, we present the first experimental demonstration of an on-chip microwave spin-plasmon interface (MSPI). The MSPI consists of an NV ensemble (NVE) coupled to a gold v-groove (VG) nanophotonic collector (Fig. 1). We numerically simulate the coupling of the NVE to a plasmonic mode (Fig. 2) and characterise the optical properties of the coupled NVE (Fig. 3). We finally detect the spin resonance by exciting the spins through the microwaves supported by the gold film and collecting the fluorescence into the VG (Fig. 4).

# RESULTS

## Design of a fluorescence collector with moderate decay rate and high coupling efficiency

We performed 3D numerical simulations of NV-VG interface and found that the VG geometry simultaneously satisfies the conditions of high coupling efficiency and moderate emission rate enhancement. In the experiment, each nanodiamond (ND) typically contained on average 400 NVs. Therefore, in the simulation we varied the dipole's position along the z-axis to study the dependence of NV emission characteristics on its spatial location, where $z = 0$ concurs with the ND centre (Fig. 2a). The calculated total decay rate values, normalised to the decay rate in bulk diamond [33], $12.8^{-1}$ ns$^{-1}$, span from 0.23 to 0.66, at the NVE peak emission wavelength of 665 nm (see Fig. 2c, black curve). All the NV centres in the ensemble were therefore expected to be subradiant with respect to NV centres in bulk diamond, favouring sensitive spin readout in the weak excitation regime.

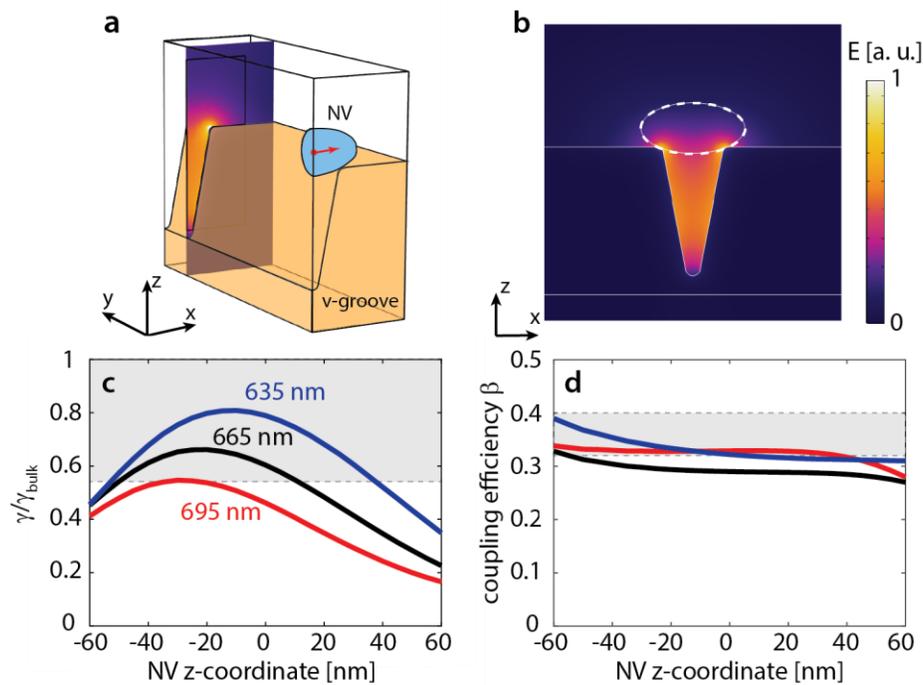

Fig. 2. Numerical simulations of NVE-VG interface: (a) E-field cross-section at a distance of 750 nm from NV. The simulation domain contains a gold VG and an NV as a dipolar emitter (red dot) embedded into a diamond ellipsoid. (b) Distribution of E-field magnitude corresponding to a fundamental channel plasmon-polariton mode excited in the VG, obtained from 2D eigenmode analysis. (c) Simulated total decay rate $\gamma$ normalised to the intrinsic rate of an NV inside a bulk diamond ($\gamma_{bulk} = 12.8^{-1}$ ns$^{-1}$). (d) Simulated coupling efficiency $\beta$ (into both collector directions) vs emitter vertical position inside an ND ($z = 0$ aligns with ND centre). Grey areas on (c) and (d) cover the range of experimentally obtained values.

The coupling efficiency of the NVE to a propagating plasmonic mode was estimated by integrating the power density in the *zx*-plane at a distance $y_0 = 0.75$ µm away from NV (see cross-section plane in Fig. 2a). The integrated power was divided by the propagation loss factor ($\exp(-y_0/L_p) \sim 0.8$) of the fundamental plasmonic mode Fig. 2b), with $L_p = 3.14$ µm being the propagation length at 665 nm. As a result, the *β*-factor was found to be around 0.3, almost independent of the NV's position along the *z*-axis (Fig. 2d, black curve). Additionally, we obtained numerical values for *β* and *γ* at 635 nm and 695 nm (Fig. 2c, d; red and blue curves), which correspond to the red and blue edges of the NVE emission spectrum, respectively (see Supplementary Fig. 6a). The coupling efficiency in the NV-VG geometry depends weakly on both the NV position inside nanodiamond and its emission wavelength thereby promising good repeatability of the results.

**Experimental measurements of fluorescence lifetime and coupling efficiency**

We fabricated an MSPI by milling VGs into a thick gold film and depositing NDs on the surface of the sample (see Methods, Sample fabrication). An image of an ND coupled to the v-groove is shown in Fig. 3a. The NVE fluorescence decay was measured by using a time-correlated single-photon counting method and the data was fitted with a gamma-weighted integral of exponential functions convolved with the instrument response function (Fig. 3b). More details can be found elsewhere [27]. The fitting-retrieved averaged lifetime was $\tau_{av}$ = 16.2 ns ($\gamma_{exp}/\gamma_{bulk}$ = 0.79) and lifetime dispersion $\delta\tau$ = 10.5 ns ($\delta\gamma/\gamma_{bulk}$ = 0.5). The range of experimental fluorescence lifetime values was visually marked by the grey area in Fig. 2c. Experimental decay rates are slightly faster than the theoretically predicted ones, the discrepancy being most likely due to the contribution of non-radiative decay paths in nanodiamond-based NVs [34].

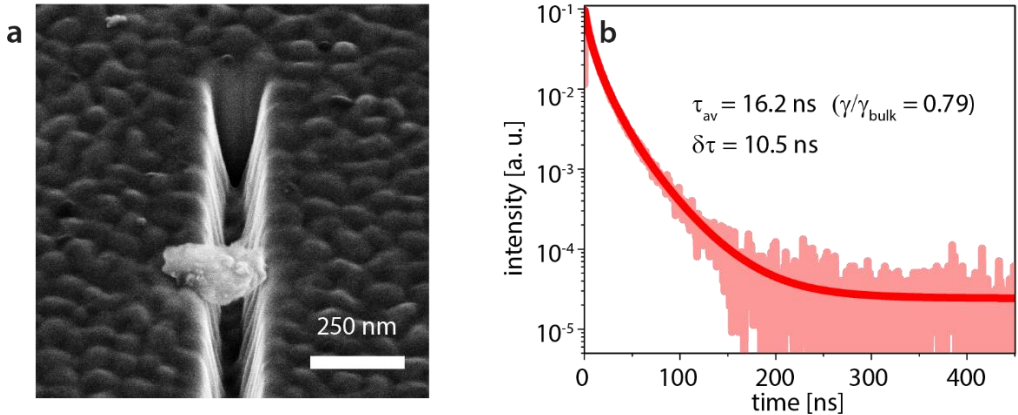

Fig. 3 Nanodiamond inside a v-groove collector: (a) SEM image of the ND coupled to a gold VG. (b) Fluorescence decay curve fitted with the gamma-distributed sum of exponential decays, average lifetime $\tau_{ave}$ of 16.2 ns (or $\gamma/\gamma_{bulk,exp}$ = 0.79 ) with a spread $\delta\tau$ of 10.5ns.

The coupling efficiency was measured by analyzing optical signals collected from the VG ends (A, B) and ND position (C), see Fig. 4a. The ND was located at distances of $l_A$ = 0.85±0.03 μm and $l_B$ = 9.85±0.03 μm from A and B. Fig. 4b shows distinct fluorescence spots corresponding to A, B, and C obtained on the CCD camera upon exciting the NVE with a 532nm focused laser beam. The brighter spot consisted of two partially overlapping peaks formed by direct emission from ND and plasmons scattered from the A end (red and pink peaks on Fig. 4c). The fainter spot corresponded to plasmon scattering at B. The fluorescence intensity at B was affected by propagation losses (transmittance ~ 0.05) and a finite NV-to-plasmon coupling (~0.2 in a single direction). Relative emission powers at A, B, C spots were retrieved from integrating pixel counts over the areas of deconvolved peaks: $P_A = 3000 \pm 50$ au, $P_B = 200 \pm 15$ au, $P_C = 14000 \pm 100$ au. Plasmon propagation length $L_p$ and NVE-VG coupling efficiency $\beta$ were calculated as: $L_p = (l_B - l_A)/\ln(P_A/P_B) \approx 3.3 \pm 0.1$ µm, $\beta = \left(P_A e^{l_A/L_p} + P_B e^{l_B/L_p}\right) / \left(P_A e^{l_A/L_p} + P_B e^{l_B/L_p} + P_C\right) \approx 0.36 \pm 0.04$.
The experimentally measured range of $\beta$ values was greyed out in Fig. 2d. Hence, we experimentally verified that the VG-coupled NVE features a coupling efficiency of nearly 0.4, which is in good agreement with the simulated results.

**Measurement of NVE electron-spin resonance through the channel plasmon-polaritons**

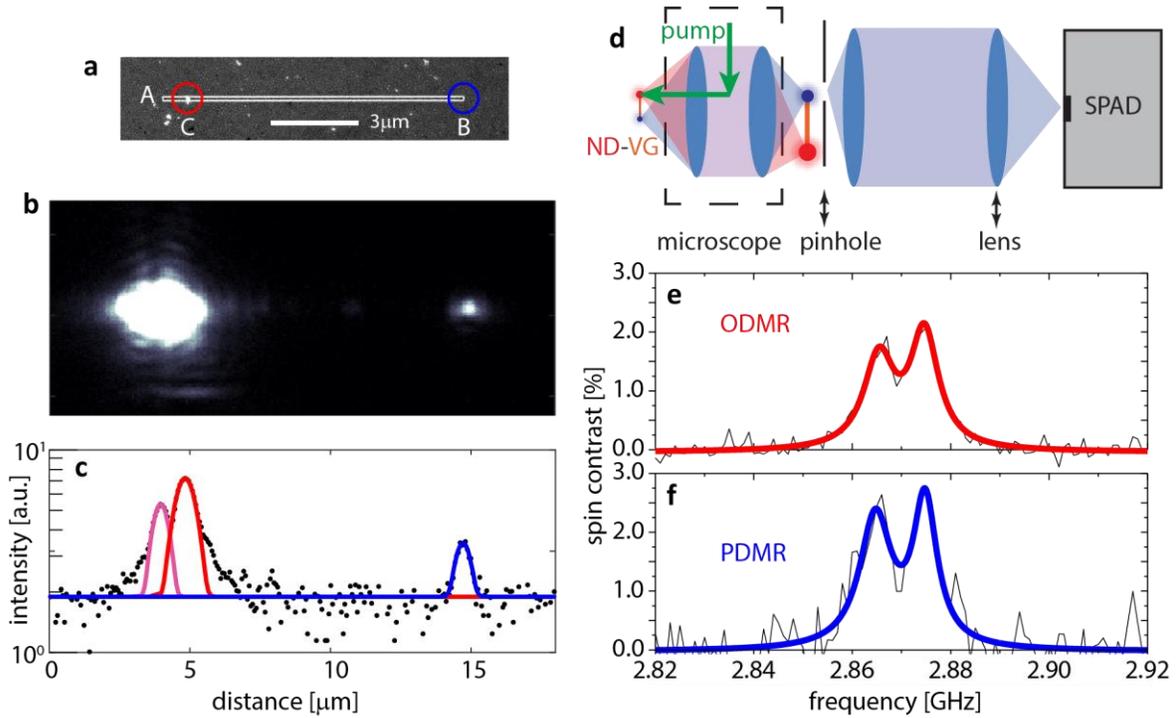

Fig. 4. NV electron spin readout via plasmons. a) SEM scan of the gold v-groove collector, 10.7 µm long with the ND outlined with a red circle. The distances between ND and the VG ends are 0.85µm (A) and 9.85µm (B). b) Photoluminescence image obtained with a CCD camera depicts merged fluorescence spots at A and C and fainter fluorescence scattering from B. c) Profile cut of the CCD image along the propagation direction with Gaussian fits of the maxima at A, B and C. d) Schematic of the experimental method for collecting the fluorescence signal from B by adjusting the pinhole of the confocal microscope. e) and f) The spin-resonance curves measured using fluorescence from spots B (optically detected magnetic resonance) and C (plasmonically detected magnetic resonance), respectively.

After characterising the NVE fluorescence, we aimed at measuring the spin resonance of the NVE. We measured the electron spin resonance by applying MW fields and detecting the fluorescence following the sequence protocol described in Methods (see Supplementary Fig. 7). The spin resonance experiments were performed at zero magnetic field by sweeping the MW frequency around 2.87 GHz, and analysing changes in fluorescence from points C (ODMR) and B (plasmonically detected magnetic resonance – PDMR, see Fig. 4e,f). Importantly, the MW excitation was delivered directly through the VG-patterned gold film. Such a metal-based design significantly simplifies fabrication, shrinks the device footprint, and allows delivery of MW excitation over a large area. The NVE optical transitions were induced by a 532 nm laser at excitation rates well below the saturation regime (see Supplementary Fig. 6b). PDMR was measured by collecting the fluorescence scattered from waveguide end B, while continuing to excite the NVE optically. For this purpose, we kept the excitation laser aligned on the ND and adjusted the pinhole and detector lens position in our homemade confocal microscope setup (see Fig. 4d). Both ODMR and PDMR signals exhibited a double peak structure with a 10 MHz inter-peak splitting, indicating presence of local strain in the nanocrystal [13]. Both of the spectra are almost identical, but the PDMR method yields about 25% higher contrast. This increased spin contrast may be due to partial filtering of the fluorescence from $NV^0$ signal which is blue-shifted compared to $NV^-$ and features no spin contrast.

The estimated single-shot spin-readout SNR and magnetic field sensitivity of the fabricated MSPI are on the order of 0.01 and 100μT/Hz$^{1/2}$. These characteristics can be further improved by reducing losses in the collector, for instance, by shortening its length. The plasmonic readout signal at the VG end can be converted to photonic low-loss modes [35–37] or sent directly to a local integrated detector, further improving the sensitivity. Additionally, MSPI can be fabricated from an epitaxial silver with an order of magnitude lower losses [38,39].

**DISCUSSION**

On-chip integration is a key factor for enabling high-efficiency and small footprint NV-based quantum devices. Fabrication complexity is one of the major stumbling blocks in achieving this goal. We are addressing this challenge by implementing a microwave spin-plasmon-microwave interface in a single material layer and one patterning step. This on-chip MSPI prototype has the potential to further advance the development of quantum sensing and quantum information applications. This device enables compact sub-micrometre-scale footprints, higher operational speeds, and reduced fabrication complexity.

Previously, the problem of on-chip large area MW excitation was addressed by constructing resonant antennae encompassing NVEs, such as a double split-ring resonator [40] and inductor[41]. Both of the works attained magnetic field deviations within 5% over the distances of 1 mm and 50 μm, respectively. Here we follow an approach with a simpler design, where NDs are brought in direct contact with the metal that carries MW. Our experiments showed that microwave signal could be delivered over a larger area (10×10 mm$^2$) without fabricating resonant structures and therefore without compromising RF bandwidth. From RF numerical simulations in COMSOL, we estimated magnetic field distribution over the sample area of 10×10 mm$^2$ (Supplementary Fig. 8). The magnitude of the field spreads with a nearly-linear gradient of about ±5% over a 1-mm-distance. Our approach enables simultaneous drive of hundreds of thousands of MSPI units with a Rabi frequency spread of 5%.

In most scenarios involving non-resonant optical pumping, it is important to suppress the residual pump power with, e.g. an optical filter. A waveguide-coupled excitation filter tailored for the NV centre fluorescence spectrum (Fig 5a) has not yet been demonstrated. Our nanophotonic collector, through its unique dispersion characteristics, offers a high transmission in the required spectral window, while strongly attenuating the pump at 532 nm. We calculated propagation lengths at different wavelengths for the VG fundamental modes (orange line in Supplementary Fig. 9), and consequently, VG transmittance at a propagation distance of 1 μm (blue line in Supplementary Fig. 9). Additionally, the VG partially attenuates the unwanted $NV^0$ fluorescence, which further improves the spin contrast.

In the future, the drop-casting method that we utilised to position the ND can be replaced by the hybrid electrothermoplasmonic nanotweezer (HENT) technique, that promises faster and more precise placement of NDs into plasmonic waveguides [42]. The HENT microfluidic dragging force appears due to a synergy between localised thermal gradients of plasmonic hotspots and externally applied a.c. electric fields. Hence, the MSPI layout comprised of a groove-patterned gold film is a particularly suitable configuration for implementing the HENT technique. Additionally, the precharacterised NDs can be controllably positioned inside the nanophotonic collectors by utilizing AFM-based pick-and-place techniques[43,44].

The metallic film employed in our experiment not only supports optical plasmons and microwave signals but can also carry a DC voltage bias. It is therefore suited to stabilise the negative $NV^-$ charge states. Band bending induced by the potential difference across an ND can stabilise the negative $NV^-$ charge state [45]. This control of the charge state is expected to increase both PDMR contrast magnitude and SNR, leading to improved spin-based sensing protocols. Additional enhancement of NVE spin-induced plasmonic readout SNR by order of magnitude can be attained by utilising the spin-to-charge state conversion and charge-state detection protocols [46].

Another opportunity resides in realising a low-temperature spin-plasmon chiral interface by employing spin-momentum locking of plasmon-polaritons [47,48]. For instance, circularly polarised optical transitions in NV centres[49] or other atom-like systems[1] can be mapped to the plasmon propagation momentum. This functionality can enable on-chip spin-induced emission routing and further exploration of novel chiral quantum architectures.

## METHODS

**Numerical simulations:**
Eigenmode 2D analysis (COMSOL 5.3a, Waves and Optics Module) was performed to investigate the fundamental v-groove mode parameters: field distribution, propagation length, and mode index. The simulation domain was a square box with an area of 1×1 µm². Scattering boundary conditions were applied on its edges, and the wavelength fixed to 665 nm, corresponding to the emission spectrum maximum. The groove's geometrical parameters were the following: opening angle - 21.6°, width - 160 nm, depth - 338 nm, radius of curvature at the bottom edge - 20 nm, radius of curvature at the side edges - 15 nm. These parameters were extracted from the SEM measurements of the fabricated VGs. Gold dielectric permittivity at 665 nm was $\varepsilon = -12.35 + 1.05i$, as obtained from the film's variable-angle spectroscopic ellipsometry characterisation (J.A. Woollam Co.; W-VASE). The supported fundamental mode index is $n_{VG} = 1.17 + 0.017i$, corresponding to a propagation length $L_P$ of 3.14 µm, mode area $A_M$ of about 0.05 µm², and respectively FOM ($L_P/A_M^{1/2}$) of 14.

3D full-wave simulations were used to estimate the total decay rate $\gamma$ and collection efficiency $\beta$. The computational domain size was 1.8×0.6×0.85 µm³ (length × width × height). It was surrounded by a perfectly matched layer (PML) shell with a thickness of 300 nm in order to suppress reflection from the boundaries. The ND was modelled as an ellipsoid with principal semi-axes of a = 135 nm, b = c = 65 nm. As a result of the ellipsoid reposing on the VG edges, the ND's centre was lifted with respect to the gold-air planar interface by ~50 nm. The choice of ND shapes and sizes was based on the obtained SEM images of the ND-VG. An NV centre was modeled as a volume current density oscillating coherently inside a 1-nm-radius sphere serving as a dipole emitter. The electric dipole orientation was chosen along the E-field of VG fundamental mode. Due to symmetry considerations, we partitioned the computational domain into four subdomains and performed simulations on one of them.

The total decay rate is assumed to be proportional to the total released power $P$ and, hence, it was estimated as $\gamma/\gamma_{bulk} = P/P_{bulk}$, where $\gamma_{bulk}$ and $P_{bulk}$ are the reference quantities of an NV in bulk diamond. $P_{bulk}$, power emitted by a dipole inside diamond environment, was analytically estimated with a standard textbook formula [50]. $P$ was retrieved from NV-VG numerical simulations as a surface integral of power flow **S** through a 3-nm-radius spherical surface encapsulating the emitting dipole and situated entirely within the ND volume $P = \int_{ND} \mathbf{S} \cdot d\mathbf{A}$. The NV-VG $\beta$-factor was calculated as $\beta = P_{VG}/P$, where P$_{VG}$ is the power emitted into the VG modes propagating in both directions, $P_{VG} = 2e^{l/L_P} \int_{A_{CS}} \mathbf{S} \cdot d\mathbf{A}$. $A_{CS}$ is a rectangular cross-section area 0.8×1.04 µm² positioned at a longitudinal distance $l = 0.75$ µm from the NV center. $L_P = 3.14$ µm the propagation length of the VG fundamental mode, was obtained from 2D mode simulations.

**Sample fabrication**

A 1×1cm$^2$ Si substrate was cleaned with a stripper Nano-Strip 2X, Cyantek (mixture of sulfuric and peroxymonosulfuric acids, hydrogen peroxide, water) for 10 minutes, then rinsed with deionised water and dried with nitrogen. Deposition was performed using an e-beam evaporator, operating at a base pressure of 5×10$^{-7}$torr. To improve the adhesion, we introduced a 5 nm thick Ti wetting layer. A 380-nm thick gold film was deposited at a rate of 2 Å/s and its thickness was characterised by profilometry measurement.

We have milled a total of 18 channel v-grooves by bombarding the gold film with a focused Ga$^+$ beam (Nova 200, FEI). The milling was performed under 30 kV accelerating voltage, with a current of 10 pA, magnification of approximately 20k. The size of a single-shot hole was measured to be 35 nm. The patterning sequence was encoded into a stream file with the specified parameters: single loop, pixel spacing – 3nm, dwell time – 150 µs, randomised sequence order.

The dimensions of fabricated v-grooves were obtained from SEM: a length of 10.70±0.05 µm and a separation distance (between v-grooves) of 5 µm. Additionally, from a FIB-cut cross-section of a similar VG, the following dimensions were measured: a groove depth of 340±10 nm, an angle of 21.6±0.5°, and a width of 160±10 nm.

NDs with an average size of 100 nm and containing nominally 400 NV centres (Adamas Nano) were dropcast onto the patterned metal surface. During the dropcasting procedure, we dissolved the initial solution (1mg/mL) 1000 times. We put a droplet of 10 µl onto the sample, covered it with a coverslip for 1 minute, after this we rinsed with DI water and nitrogen dried. Before depositing NDs the patterned gold film was covered with a poly-allyalamine hydrochloride (PAH) layer to improve adhesion of individual nanocrystals to the gold surface. The SEM study proved that several nanodiamonds were placed into or very close to v-groove.

**NVE optical characterization**

The optical signals were collected and analysed using a custom-made confocal microscope setup based on a Nikon Ti-U microscope body and equipped with a Nikon air objective (NA 0.9), a PI-561 objective piezo-stage driven by a E-712 controller (Physik Instrumente), a 532 nm CW laser (SLOC, GL532NA-200), a 514 nm picosecond laser (BDL-514-SMNi, Becker & Hickl), SPAD detector (SPCM-AQRH, Excelitas), an optical spectrometer (Ocean Optics QE65000) and a CCD camera (ATIK 414EX). The pump reflection was filtered by a long-pass 550 nm filter with OD6. Microwaves were generated by an Agilent E8254A generator, modulated by a MiniCircuits ZASW-2-50DR+ switch, and amplified with 16 W Mini-Circuits ZHL-16W-43 RF amplifier. The cable from the MW amplifier was soldered onto the sample's 380-nm-thick gold film.

**Data availability**

The datasets generated during and/or analysed during the current study are available from the corresponding author on reasonable request.

**ACKNOWLEDGEMENTS**

The authors acknowledge R. Cui and S. Choudhury for the help with v-groove fabrication, and A. V. Akimov, V. Vorobyov for useful discussions. We additionally would like to thank Samuel Peana for help with manuscript preparation. This work was partially supported by the U.S. Department of Energy, Office of Basic Energy Sciences, Division of Materials Sciences and Engineering under Award DE-SC0017717 (S. I. Bogdanov), and the Office of Naval Research (ONR) DURIP Grant N00014-16-1-2767 (equipment grant used to purchase the scanning confocal microscope, lasers, detectors, used in this work). A. V. Kildishev acknowledges the DARPA/DSO Extreme Optics and Imaging (EXTREME) Program, Award HR00111720032 (numerical modelling and simulations).

**AUTHOR CONTRIBUTIONS**

M. Y. Shalaginov and S. I. Bogdanov conceived the project. M. Y. Shalaginov and A. V. Kildishev carried out the numerical simulations. M. Y. Shalaginov performed sample fabrication and characterization. S. I. Bogdanov, M. Y. Shalaginov and A. S. Lagutchev built the optical experimental setup for measuring plasmonically detected magnetic resonance and photoluminescence characterization. M. Y. Shalaginov and S. I. Bogdanov collected and analysed experimental data. A. V. Kildishev, A. Boltasseva and V. M. Shalaev supervised the project. All authors discussed results and contributed to the writing of the manuscript.

**COMPETING INTERESTS**

The authors declare no competing financial interests.

# SUPPLEMENTARY INFORMATION

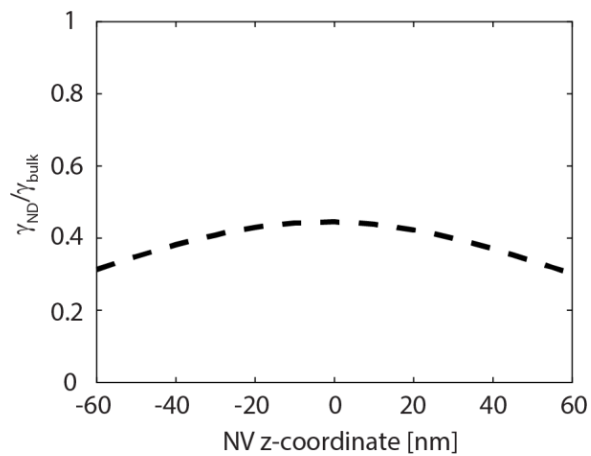

Fig. 5. The ratio of the total decay rate of an NV inside an ND immersed in air ($\gamma_{ND}$) to $\gamma_{bulk}$.

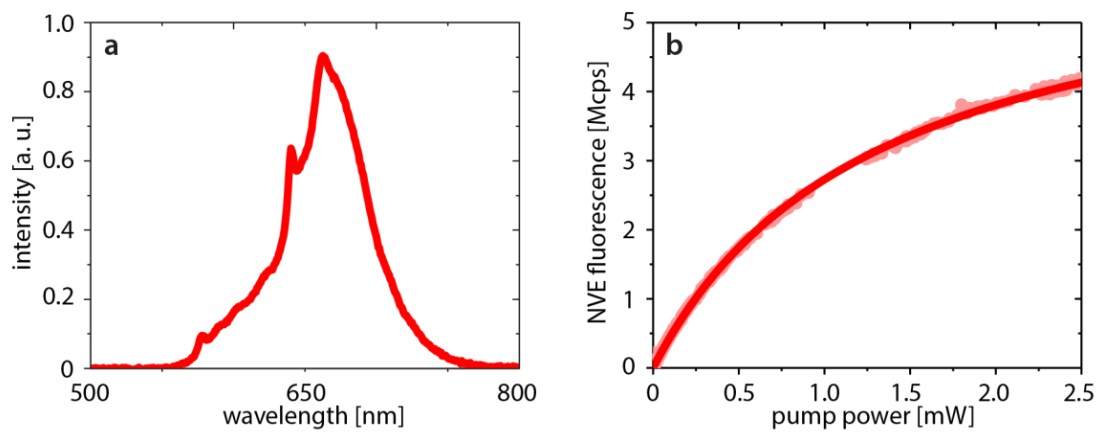

Fig. 6. (a) Emission spectrum of NVE centre excited with a 532-nm laser. (b) Saturation dependence of NVE fluorescence vs laser pump power extrapolated saturated fluorescence is 6.4 Mcps, saturation pump power - 1.3 mW. In spin readout measurements the excitation power was 15 μW.

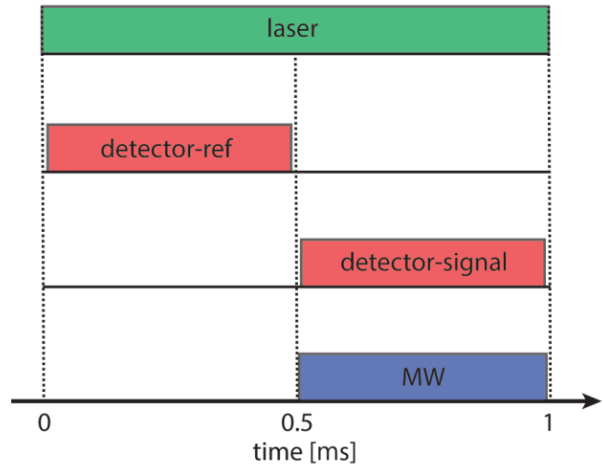

Fig. 7. A protocol of spin-resonance measurement cycle: continuous excitation by 532 nm (green); two counters (red) connected to a SPAD through a t-adapter: reference (MW off) and signal (MW on) time windows: 0.03 - 0.47 ms and 0.53 - 0.97 ms, respectively. MW excitation (blue), modulated by a switch, was applied in a time period of 0.52 – 0.98 ms. The MW frequency was swept in the range from 2.82 to 2.92 GHz with a step of 0.01GHz. At each frequency step the measurement cycle was repeated 10000 and 60000 times for ND (C) and VG far end (B) signals, correspondingly.

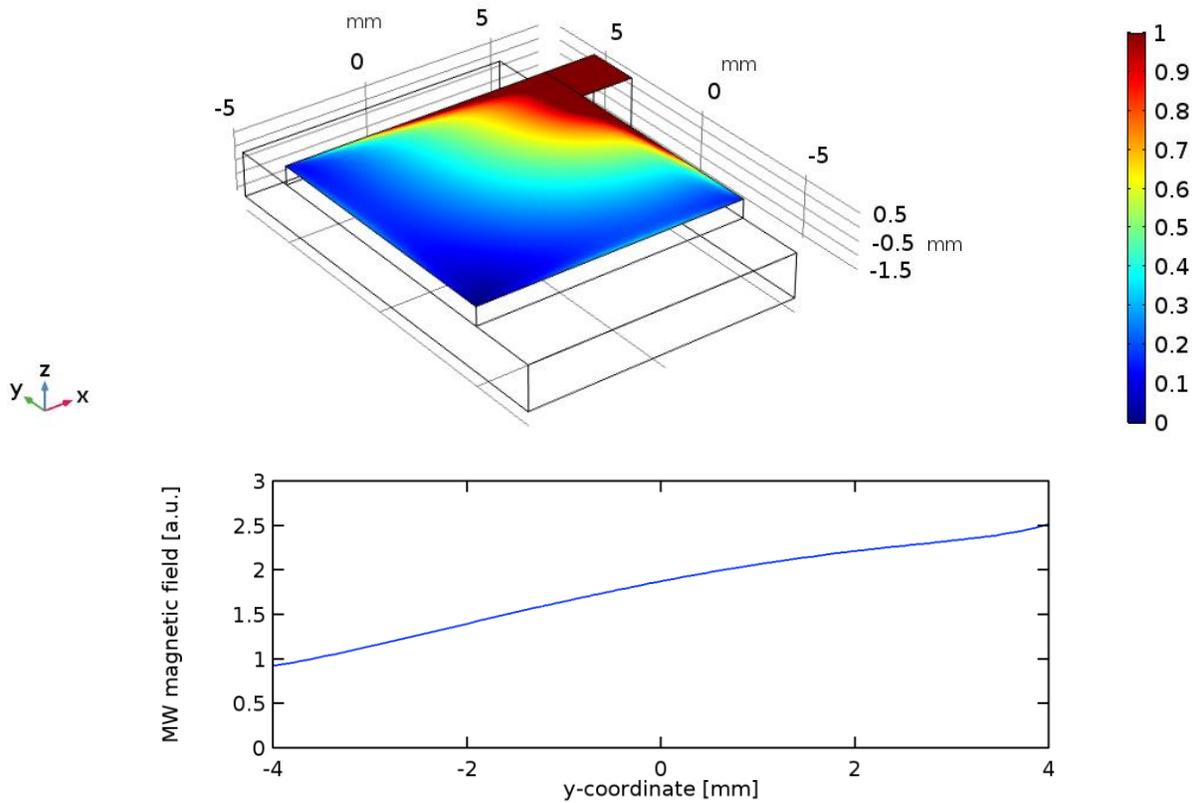

Fig. 8. Simulated MW (2.87GHz) H-field distribution on 10×10 mm$^2$ sample surface. The estimated gradient of the field magnitude gradient is about ±5% per 1mm in arbitrary direction.

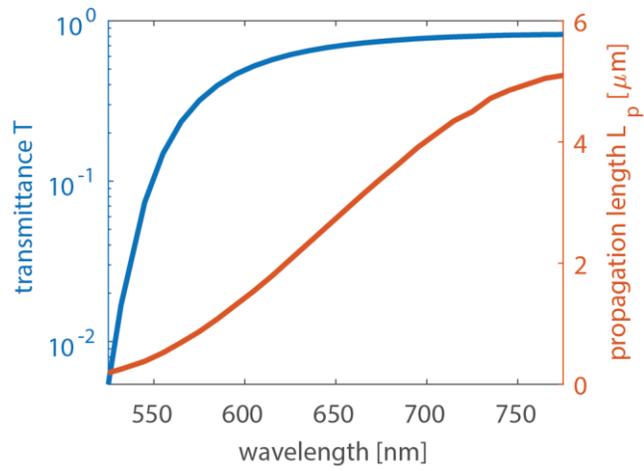

Fig. 9. Calculated wavelength dependencies of the VG fundamental mode propagation length (orange) and collector transmittance at the propagation distance of 1 μm (blue). Demonstrated VG can be used as a long pass filter for cutting off the pump at 532 nm.